\documentclass[epj]{svjour}
           
\usepackage{isolatin1,epsfig,graphics}
\begin{document}
\title{Distributed Computing Concepts in D\O}
\author{Daniel Wicke
}
\institute{Bergische Universit\"at Wuppertal\\
           Gau\ss{}str.~20\\
           D-42097 Wuppertal\\
          \email{Daniel.Wicke@physik.uni-wuppertal.de}
}

\newcommand{\vektor}[2]
{{#1 \choose #2}}
\newcommand{\SU}{SU}
\newcommand{\SO}{SO}
\newcommand{\U}{U}

\newcommand{\hc}{\mbox{h.c.}}
\newcommand{\Tr}{\mbox{Tr}}

\newcommand{\PC}{Potenzreihenkorrektur}
\newcommand{\PCen}{Potenzreihenkorrekturen}

\newcommand{\Slash}[1]{\mbox{\it #1\hspace{-0.6em}\slash}}

\newlength{\ziffer}
\newcommand{\0}{\settowidth{\ziffer}{0}\hspace*{\ziffer}}

\newcommand{\TeV}{\,\mbox{Te\kern-0.2exV}}
\newcommand{\GeV}{\,\mbox{Ge\kern-0.2exV}}
\newcommand{\mGeV}{\,\mathrm{Ge\kern-0.2exV}}
\newcommand{\MeV}{\,\mbox{Me\kern-0.2exV}}
\newcommand{\keV}{\,\mbox{ke\kern-0.2exV}}
\newcommand{\eV}{\,\mbox{e\kern-0.2exV}}
\newcommand{\km}{\,\mbox{km}}
\newcommand{\m}{\,\mbox{m}}
\newcommand{\cm}{\,\mbox{cm}}
\newcommand{\mm}{\,\mbox{mm}}
\newcommand{\um}{\,\mbox{$\mathrm\mu$m}}
\newcommand{\fm}{\,\mbox{fm}}
\newcommand{\us}{\,\mbox{$\mathrm\mu$s}}
\newcommand{\hz}{\,\mbox{Hz}}
\newcommand{\ipb}{\,\mbox{pb}^{-1}}
\newcommand{\ifb}{\,\mbox{fb}^{-1}}
\newcommand{\pb}{\mbox{pb}}
\newcommand{\mb}{\mbox{mb}}

\newcommand{\bea}{\pagebreak[3]\begin{samepage}\begin{eqnarray}}
\newcommand{\eea}{\end{eqnarray}\end{samepage}\pagebreak[3]}
\newcommand{\beq}{\begin{equation}}
\newcommand{\eeq}{\end{equation}}

\newcommand{\Bildzentriert}[3]
{  \begin{figure}[#2]
   \begin{center}
   \input #1
   \end{center}
   \caption{#3}
   \label{#1}
   \end{figure}
}

\newcommand{\Bild}[3]
{  \begin{figure}[#2]
   \input #1
   \caption{#3}
   \label{#1}
   \end{figure}
}

\newcommand{\C}{{C\hspace{-0.65em}I}}
\newcommand{\R}{{I\hspace{-0.35em}R}}
\newcommand{\N}{{I\hspace{-0.35em}N}}

\newcommand{\sm}{Standardmodell}
\newcommand{\ww}{Wechselwirkung}
\newcommand{\ee}{$e^+e^-$}
\newcommand{\as}{$\alpha_s$}
\newcommand{\Bmax}{B_{\mathrm{max}}}
\newcommand{\Bmin}{B_{\mathrm{min}}}
\newcommand{\Bsum}{B_{\mathrm{sum}}}
\newcommand{\Bdiff}{B_{\mathrm{diff}}}
\newcommand{\Mhigh}{M^2_{\mathrm{h}}/E^2_{\mathrm{vis}}}
\newcommand{\Mlow}{M^2_{\mathrm{l}}/E^2_{\mathrm{vis}}}
\newcommand{\Mhighp}{{M^{2 }_{(p)\mathrm{h}}}/E^2_{\mathrm{vis}}}
\newcommand{\Mlowp}{{M^{2}_{(p)\mathrm{l}}}/E^2_{\mathrm{vis}}}
\newcommand{\MhighE}{{M^{2}_{(E)\mathrm{h}}}/E^2_{\mathrm{vis}}}
\newcommand{\MlowE}{{M^{2}_{(E)\mathrm{l}}}/E^2_{\mathrm{vis}}}
\newcommand{\Mdiff}{M^2_{\mathrm{diff}}/E^2_{\mathrm{vis}}}
\newcommand{\durham}{{Durham}}

\newcommand{\ecm}{E_{\mathrm{cm}}}

\newcommand{\asb}{$\alpha_0$}

\newcommand{\eps}{\varepsilon}
\newcommand{\abb}{Fig.~\ref}
\newcommand{\fig}{\abb}
\newcommand{\tab}{Tab.~\ref}
\newcommand{\gl}[1]{Gl.~(\ref{#1})}
\newcommand{\eq}[1]{Eq.~(\ref{#1})}

\newcommand{\oas}{$\cal O$($\alpha_s^2$)}
\newcommand{\oass}{$\cal O$($\alpha_s^3$)}

\newcommand{\sprime}{{\sc Sprime}}
\newcommand{\pythia}{{\sc Pythia}}
\newcommand{\jetset}{{\sc Jetset}}
\newcommand{\ariadne}{{\sc Ariadne}}
\newcommand{\herwig}{{\sc Herwig}}
\newcommand{\excalibur}{{\sc Excalibur}}
\newcommand{\minuit}{{\sc Minuit}}
\newcommand{\zfitter}{{\sc Zfitter}}
\newcommand{\event}{{\sc Event}}

\newcommand{\lep}{{\sc LEP}}
\newcommand{\delphi}{{\sc Delphi}}
\newcommand{\alephh}{{\sc Aleph}}
\newcommand{\opal}{{\sc Opal}}
\newcommand{\ldrei}{{\sc L3}}
\newcommand{\sld}{{\sc SLD}}
\newcommand{\delana}{{\sc Delana}}
\newcommand{\dstana}{{\sc Dstana}}
\newcommand{\delsim}{{\sc Delsim}}
\newcommand{\Mini}{{\sc Mini}}
\newcommand{\phwmini}{{\sc PHWMini}}
\newcommand{\cargo}{{\sc Cargo}}
\newcommand{\jade}{{\sc Jade}}

\newcommand{\tsppm}{\hspace{\tabcolsep}$\pm$\hspace{\tabcolsep}}

\newenvironment{scaledlist}[0]{
                          \begin{flushleft}
                          \begin{list}{{$\bullet$}}{\setlength{\itemsep}{2ex plus0.2ex}
                          \setlength{\parsep}{0ex plus0.2ex}
                          \setlength{\labelwidth}{2em}}
                         }
                         {
                          \end{list}
                          \end{flushleft}
                         }


\abstract{                               
The D\O\ experiment faces many challenges enabling access to large datasets
for physicists on four            continents. The new concepts for distributed
large scale computing implemented in D\O\ aim for an optimal use of the
available computing resources while minimising the person-power needed for
operation. The real live test of these concepts is of special interest for the
LHC Computing GRID, LCG, which follows a similar strategy.
}
\maketitle

\section{Introduction}
\label{sec:intro}
Some of the most interesting events in $p\bar{p}$ collisions at the
Tevatron are very difficult to distinguish from the overwhelming QCD background.
Top and Higgs particles are moreover very rare. The Tevatron experiments
D\O\ and CDF therefore record large amount of data for later analysis and
detailed studies. While taking data each experiment records roughly 500\,GB of raw data 
per day. Reconstructing these events adds another 1.1\,TB/day in D\O.
Within the last two years around 350\,TB have been accumulated.
Providing this data volume for physics analysis performed by more than 
100 people is one of the challenges D\O\ faces.

With this amount of data providing the necessary overall IO rate is a difficult
task. As only a fraction of the data can be stored on disks, tape mounting
leads to major dead-times. D\O\ follows a combined concept of locally 
optimising the resource usage and distributing the data globally.
For an international collaboration like D\O\ with around 50\% of the
collaborators stemming from non-US institutes the second step is
of special importance to
provide easy data access not only for those resident near Fermilab but also
to those working remotely often on another continent.

In addition, to analyse these data,
sufficiently many simulated events need to be
produced for selection studies and the estimation of detector effects. 
To fulfil the requests the production is distributed to many sites. 
Production chains with ever changing versions and parameters are
however complicated to handle. 
To ease the production an 
automatic handling of large batches of jobs was developed.

The mentioned concepts to meet the outlined requirements are discussed 
in the following.

\section{Management of large batches of jobs}
\label{sec:runjob}
\begin{figure}[b]
\centerline{\epsfig{file=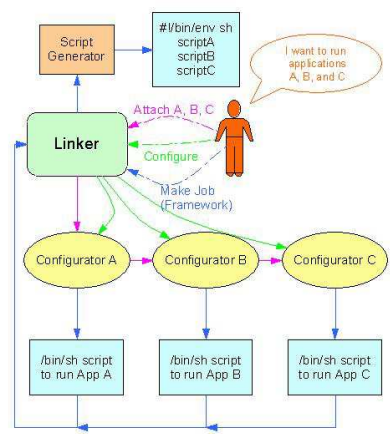,width=0.3\textwidth}}
\caption{\label{fig_runjob}{\tt Runjob} separates the details of how to call
  individual programs (A, B and C) from the operational details of which
  programs (or program versions) to combine.
}
\end{figure}
To reduce the person-power needed  for Monte Carlo
production with its ever changing versions and parameter settings,
D0 developed the work-flow management system  {\tt Runjob}
\cite{runjob01,runjob03}. A work which is continued 
in collaboration with CMS~\cite{shahkar}.

It automates the linking of several processing tasks 
into a single job based on a job
description and allows to separate the configuration of the individual programs
from the definition of the work-flow (\fig{fig_runjob}). 
Thus the individual steps of Monte Carlo
production, like event generation, detector simulation and reconstruction,
can be configured by the corresponding experts. Those performing the
production can in turn concentrate on the work-flow, i.e. which program versions
to combine.

Furthermore, the coherent description of the production work-flow allows
the preservation of the meta-data describing what actually was run during the
production. 
D\O\ stores these meta-data along with the produced simulated data in the
SAM-system described below.

Recently the work-flow description was extended to allow for job
parallelisation. This way a single production can be run in many parallel
jobs  at a given site. {\tt Runjob} takes care about the necessary
initialisation and termination tasks, e.g. the combination of results of those many
parallel jobs.

\section{Optimised use of local resources}
\label{sec:sam}
Beside person-power the optimised use of computing resources is important to
fully exploit the physics potentials of D\O. With the given amount simulated
and real data only a fraction of the available information can be kept on
disks. All other data are stored within tape robots.

To avoid inefficiencies due to tape mounting the access to data needs to be
optimised. D\O\ has developed a data management system (named Sequential Access
through Meta-data, SAM \cite{d0note3464}) 
which exploits that the order of the stored physics
events is of no interest for the analyses: Instead of looping over a given
list of files the user requests a dataset. The order in which the files
belonging to this dataset are processed is optimised to minimise the 
overall number of tape mounts.
The definition of a dataset can be performed by the individual users
based on the meta-data which describe the content of the files.

SAM also includes user-level bookkeeping and is a central component
of the D\O\ software environment. It has recently been adopted by CDF.

\section{Worldwide distribution of data}
\label{sec:rac}
\begin{figure}[bt]
\epsfig{file=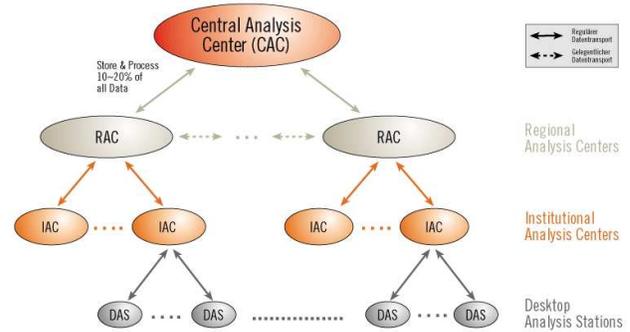,width=0.5\textwidth}
\caption{\label{fig_ractree} Data distribution scheme within D\O. 
The distribution of data is done in a tree structure from the central
repository (CAC) to regional analysis centres (RACs), further to the
institutional centres (IACs) and finally to the physicists desktop (DAS).
}
\end{figure}
Even after optimising local resource utilisation and tape mounting, the
access to the data is a bottleneck. To further improve the accessibility
of data especially for physics analysis in summer 2002 a scheme for data
distribution was outlined \cite{d0note3984} (\fig{fig_ractree}).
Beside the central data repository, which holds all D\O-data, 
several regional centres should hold copies of the data used for analysis
and provide them to users of their region.

These regional centres in addition serve the institutions in
that region such that collaborators in that institutes can develop and
test their analysis at their local computer cluster or even their desktops.

With the tree structure D\O\ hopes to minimise the necessary amount of data
copies over long distances. Beside storage such regional centres should
provide a reasonable amount of CPU power to analyse the stored data.

\subsection{Prototype}
\label{subsec:racproto}
To test the concept of regional centres which serve the associated
institutes, a prototype was set up at the German Grid
Computing Centre in Karlsruhe (GridKa) \cite{gridka} 
with the five (now six) German institutes
associated.

With this prototype the main parameters of the system should be tested:
The required network bandwidth and the manpower to run the regional
centre. Moreover shortcomings of the D\O\ software should be discovered.

GridKa was established in 2002 and is rapidly growing.
The centre is currently shared by 8 HEP experiments. Its (at the time) 
roughly 180 compute
nodes are set up with NFS-shared user home directories and NFS-shared
experiment specific disk areas. No experiment specific code can be set up
on the compute nodes. To allow for pre-Grid usage of the system, each
experiment has a so called software server, on which experiment specific
software can be installed and which serves for user login.
These specifications are quite different from the
setup used on the systems at FNAL, which are exclusively used by D\O.

Unfortunately, the required changes to the D\O-software couldn't, 
in all cases, be implemented in a site independent 
way, such that an adaption of each version and 
to each computing cluster is still necessary.
To avoid these tasks in the future a standard D\O\ computing environment needs
to be defined which is flexible enough to deal with all possible cluster
configurations. 

The network bandwidth to Fermilab is sufficient to continuously download the
most condensed data format (Thumbnails). When requesting large datasets in one
go transport rates of around 3.3MB/s (8TB/day) are observed.

All changes required for doing D\O\ analysis  at GridKa 
were available by end of 2002, with the exception of luminosity access. 
During January several German collaborators used the additional resources to
finish their analysis for Moriond in time. With this successful end-user test
the value of regional centres was finally proved.

The experiences with this prototype show, that beside the technical
achievements, it is of great advantage
for the users to work on a centre in their own time-zone, where user problems
can be solved by the operators during the usual working hours.

\section{A GRID for D\O}
\label{sec:grid}
\begin{figure}[b]
\epsfig{file=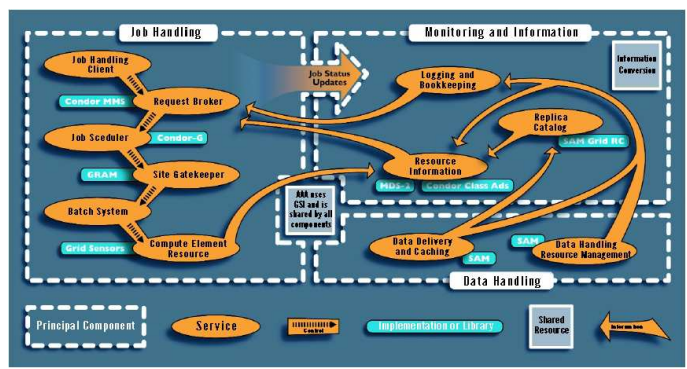,width=0.5\textwidth}
\caption{\label{fig_samgrid} Components diagram of the SAM Grid project.
}
\end{figure}
While the regional analysis centre in Germany has proven its
value, the effort that needs to be taken by the user is still large. The
analysis code needs to be copied manually to GridKa and needs to be
recompiled on the head node before it can be run. This procedure currently
needs to be repeated for each site that shall be used.
Moreover the user needs to track the availability and performance at different
sites in order to make even a reasonable choice about where to run his/her project.

To allow for an automatic dynamic adaption to the actual situation GRID tools
are required. The JIM project 
\cite{rod_beauty2002,Terekhov:2003we,Baranovski:2003wg,Baranovski:2003is} 
is an integration of existing Globus
\cite{globus} tools with the SAM system \cite{d0note3464} 
used by both D\O\ and CDF. JIM uses
Condor \cite{condor-flock} as its resource broker (\fig{fig_samgrid}). 

An initial version of JIM has been installed at several sites within D\O\
including the GridKa cluster. First experiences with physics analyses in a
GRID like environment are expected in the near future.

\section{Summary}
\label{sec:summary}
To maximise the physics output D\O\ aims to optimise the use of its resources
while reducing the person-power needed to maintain operation. 

The data management system SAM provides user-level bookkeeping 
and optimises the use of storage resources.
World wide distribution of data through regional centres to individual
institutions adds additional resources for standard tasks
like physics analyses, Monte Carlo production or data (re)processing.
JIM integrates of Globus and Condor based GRID tools to reach a
coherent access to the globally distributed resources and to allow for global
optimisation of their usage.
{\tt Runjob} eases the handling of large batches of jobs.

With these concepts D\O\ profits from additional, better exploited
resources and reduced operation tasks.
At the same time many of the distributed computing concepts foreseen for LCG
\cite{lcg} are tested in a real live environment, which is of strong interest
for the preparation of the LHC experiments. 
\bibliographystyle{unsrtnewnt}
\bibliography{Grid}

\begin{thebibliography}{10}

\bibitem{runjob01}
G.~Graham and D.~Evans.
\newblock {\em CHEP Proceedings} (2001) .

\bibitem{runjob03}
G.~Graham, D.~Evans, and I.~Bertram.
\newblock {\em CHEP Proceedings} (2003) .
\newblock ArXiv:cs.dc/0305063.

\bibitem{shahkar}
{\tt http://www.uscms.org/s\&c/testbed/Tiger/ SHAHKAR/Shahkar.htm}.

\bibitem{d0note3464}
Mike Diesburg et~al.
\newblock D\O-Note 3464, 1998.

\bibitem{d0note3984}
I.~Bertram et~al.
\newblock D\O-Note 3984, 2002.

\bibitem{gridka}
Grid Computing Centre Karlsruhe (GridKa), {\tt http://www.gridka.de/}.

\bibitem{rod_beauty2002}
Rod Walker.
\newblock Beauty 02, 2002.

\bibitem{Terekhov:2003we}
I.~Terekhov.
\newblock {\em Nucl. Instrum. Meth.} {\bf A502}(2003)  402--406.

\bibitem{Baranovski:2003wg}
A.~Baranovski et~al.
\newblock {\em Nucl. Instrum. Meth.} {\bf A502}(2003)  423--425.

\bibitem{Baranovski:2003is}
A.~Baranovski et~al.
\newblock  (2003) .
\newblock ArXiv:cs.dc/0307007.

\bibitem{globus}
I.~Foster and C.~Kesselman.
\newblock {\em Intl J. Supercomputer Applications} {\bf 11(2)}(1997)  115--128.

\bibitem{condor-flock}
D.H.J. Epema, M.~Livny, R.~van Dantzig, X.~Evers, and J.~Pruyne.
\newblock {\em Future Generation Computer Systems} {\bf 12}(1996)  53--65.

\bibitem{lcg}
LHC Computing Grid Project, {\tt http://cern.ch/lcg/}.

\end{thebibliography}
\end{document}